\documentclass[10pt, a4paper, conference, compsocconf]{IEEEtran}
\pdfoutput=1

\usepackage{tabularx}
\usepackage{url}
\usepackage{bbding}
\usepackage{ifpdf}
\ifpdf
    \usepackage{graphicx}
\else
    \usepackage[dvips]{graphicx}
\fi
\usepackage{tabularx}
\usepackage{booktabs}
\usepackage[utf8]{inputenc}
\usepackage{xcolor}
\usepackage{listings}
\newcommand{\insertprojectname}{ANANAS}

\begin{document}
%
\title{ANANAS -- A Framework For Analyzing Android Applications}


\author{
    \IEEEauthorblockN{Thomas Eder, Michael Rodler, Dieter Vymazal, Markus Zeilinger}
    \thanks{This is a preprint version of the paper accepted for publication at the \textit{First Int. Workshop on Emerging Cyberthreats and Countermeasures}, September 2-6, 2013, Regensburg, Germany.}
    \IEEEauthorblockA{Department Secure Information Systems\\
University of Applied Sciences Upper Austria\\
\{thomas.eder, michael.rodler\}@students.fh-hagenberg.at\\
\{dieter.vymazal, markus.zeilinger\}@fh-hagenberg.at}
}

\maketitle

\begin{abstract}
  Android is an open software platform for mobile devices with a large market share in
the smartphone sector. The openness of the system as well as its wide adoption lead
to an increasing amount of malware developed for this platform.
\insertprojectname\ is an expandable and modular framework for analyzing Android
applications. It takes care of common needs for dynamic malware analysis and provides
an interface for the development of plugins. Adaptability and expandability have been
main design goals during the development process.
An abstraction layer for simple user interaction and phone event simulation is also 
part of the framework. It allows an analyst to script the required user 
simulation or phone events on demand or adjust the simulation to his needs.
Six plugins have been developed for \insertprojectname. They represent 
well known techniques for malware analysis, such as system call hooking and network
traffic analysis. The focus clearly lies on dynamic analysis, as 
five of the six plugins are dynamic analysis methods.

\end{abstract}


%
\IEEEpeerreviewmaketitle

\section{Introduction}
Android is an operating system and open software platform for mobile devices based on the linux kernel.
Since its first appearance in 2008, it became 
a big success story. In the third quarter of 2012, 72.4\% of all mobile devices sold to 
end users were powered by Android \cite{gartner1}. This makes it the most installed OS on recently
sold mobile devices \cite{idc1}. The currently most used Android versions are, 2.3 (\textit{Gingerbread}) 
with a share of 39.8\%, and the versions 4.0 (\textit{Ice Cream Sandwich}), 4.1 and 4.2 (\textit{Jelly Bean}) 
with a share of 50\% \cite{androiddashboards}.

The increasing number of smartphones based on Android and the openness of the system, 
led to an increase of malware developed for the platform. In their threat report for the 
fourth quarter of 2012 \cite{mcafee-threatrepq42012}, McAfee mentions
that their ``mobile malware zoo'' accomodates a total of 36,669 samples by the end of 2012.
95\% of these samples were gathered in 2012. McAfee also observed that the growth of 
mobile malware almost doubled in each of the last two quarters of 2012. 
From all mobile malware samples that have been gathered, 97\% are targeting Android.
Android malware was even used to carry out targeted attacks, as Kaspersky 
reported in a security alert \cite{kasperskytargeted}.
This increase of malicious applications targeting the Android platform and its users, 
needs to be addressed urgently. To keep up with this huge growth rate of malware targeting 
Android, tools for automated and semi-automated malware analysis are much needed.

In this paper, \insertprojectname, a framework for \textbf{An}alyzing \textbf{An}droid 
\textbf{A}pplication\textbf{s} (APKs), focused on automated static and dynamic malware 
analysis, is presented. While other projects aim at developing different methods and tools
for Android malware analysis, \insertprojectname\ is a framework, which allows the 
implementation and integration of several different analysis methods into one powerful 
platform. It was developed with the following goals in mind:
\begin{itemize}
    \item To provide a core, which takes care of common needs for 
    dynamic analysis, such as starting a clean environment for each analysis.
    \item To be useable for all Android versions.
    \item To provide adjustable user interaction and phone event simulation.
    \item To allow the integration of plugins for static or dynamic analysis, which can be called at different stages of the analysis process.
    \item Results should be available in several formats at different levels of detail.
    \item To be able to tailor the framework to specific needs by providing high configurability.
\end{itemize}

By providing the \insertprojectname\ framework to the community, we hope to ease and
facilitate the development of new and improved malware analysis methods. Together with 
static and dynamic analysis plugins, \insertprojectname\ becomes a comprehensive 
tool for analyzing Android applications.

The remainder of this paper is structured as follows:
In section \ref{sec:relatedwork} we take a look at related projects. 
The architecture of \insertprojectname\ is described in section \ref{sec:architecture} and 
section \ref{sec:core} explains the components in greater detail. 
In section \ref{sec:plugins}, several plugins developed for the framework are 
presented. To complete this paper, in section \ref{sec:evaluation} results of an experimental
evaluation are shown.

\section{Related Work}
\label{sec:relatedwork}
A popular tool for dynamic analysis of Windows applications is 
Anubis \cite{anubis}. 
Recently it was extended to allow the analysis of Android applications (codename Andrubis) \cite{andrubispost}.
It leverages several existing tools, like 
DroidBox \cite{droidbox},
TaintDroid \cite{taintdroidpaper},
apktool \cite{apktool} and
androguard \cite{androguard} 
for static and dynamic analysis of Android applications (APKs). 
Unfortunately, not much about its inner working or architecture is public. 
However, a comparison between Andrubis and \insertprojectname\ is drawn
in the experimental evaluation section.

Another tool similar to \insertprojectname\ and Anubis is Mobile-Sandbox \cite{mobilesandbox}.
It combines static and dynamic analysis techniques for analyzing Android applications. 
Unfortunately, there were no dynamic analysis results publicly available at the time of writing.

DroidScope \cite{droidscope} is a dynamic malware analysis tool that 
is built on top of QEMU and relies on introspection of an emulated
system running the application in question. It is able to reconstruct the OS-level view (e.g. system 
calls) and the Java-level view (e.g. instructions in the Dalvik VM, Android's Java Virtual Machine). 
It is similar to \insertprojectname\ in the way that it uses dynamic
analysis techniques and exports APIs, which can be used to develop plugins.

Recently, Rastogi et al. published their work on AppsPlayground \cite{appsplayground}, 
which is a framework for automated dynamic security analysis of Android applications.
Among other things, they focus on detection evasion and automated exploration
techniques for increased coverage of the applications code and
therefore to trigger the malicious behaviour of an application.
The detection/analysis techniques of AppsPlayground are similar to those of
\insertprojectname\ but the approaches used for the implementation are different.
\insertprojectname\ for example avoids intrusive changes to the
code of the original Android system to be able to adopt new 
Android versions more quickly. 
In contrast, AppsPlayground uses TaintDroid, which modifies the original Android version heavily.

\section{\insertprojectname\ Architecture}
\label{sec:architecture}

\begin{figure}
    \centering
    \includegraphics[height=0.29\textheight]{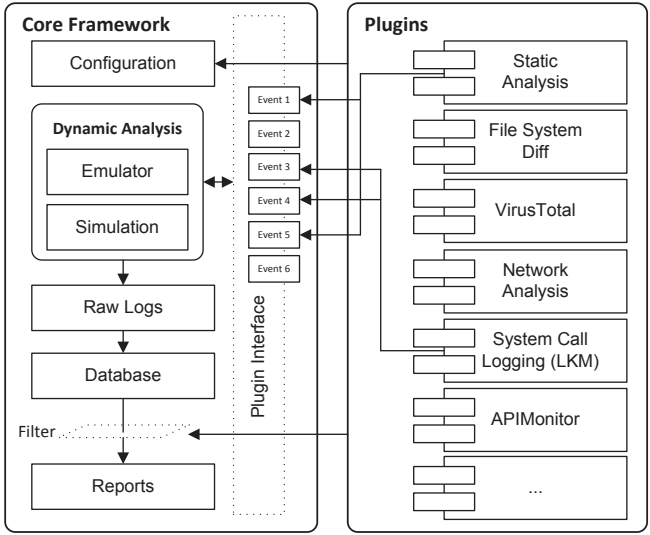}
    \caption{ANANAS Architecture}
    \label{fig:architecture}
\end{figure}

Figure \ref{fig:architecture} shows the basic architecture of \insertprojectname. 
It is composed of the core framework (written in Python) and several analysis plugins, 
which implement different analysis methods. The configuration system is shared between 
the framework and the plugins, which allows plugins to alter the configuration of the 
framework. The plugins can register to several events that are raised within the 
framework to run their code. An analysis run yields raw logfiles that are produced 
by the plugins and saved into a database. To generate a readable report, the framework 
uses filters defined by the plugins.

The framework provides common services that are needed for
the dynamic analysis. One requirement of dynamic malware analysis is that a clean, 
emulated environment is initialized for every analysis run. To emulate a smartphone, 
the emulator, which is shipped together with the Android SDK, is used.
The framework itself is independent of the used Android version.

The analysis of Android malware differs from the analysis of Windows 
based malware because most Android malware is only triggered by certain
phone events or user interaction. It is therefore crucial to
simulate interaction with the emulated smartphone during the analysis phase.
As the malicious behaviour of different malware families is triggered by
different events, the simulation has to be adjustable.
\insertprojectname\ achieves this by introducing a scripting language
for user and phone event simulation.

The modularity and extendability of \insertprojectname\ is accomplished
through the implementation of an event-based plugin system.
Individual analysis methods are implemented as plugins, which can register to several events that are 
specified within the core framework. Every time a certain event is raised, methods for all plugins registered to 
this event are called by the core framework. Currently, six plugins are implemented, which
are described in greater detail in section \ref{sec:plugins}.

Plugins can supply custom filtering mechanisms for further reducing the verbosity
of the report when searching for very specific behaviour within an Android app. The 
filters follow a blacklist approach (exceptions can be defined) and are applied during report generation.

The result of every analysis is a report containing general information about the 
Android application and the results of the different plugins. 
The results of each plugin pass through several different stages and formats. 
Primarily, each plugin saves its results to a raw log file (JSON formatted). This 
log file offers the most detailed view on the analysis run. By saving the results into 
a database in the next step, an analyst gets the chance to use SQL queries for comparing results and 
generating statistical data. Finally, the plugin specific filters are applied to
generate a condensed report in XML format, which can be used for a first examination.

An overall design goal during the development of \insertprojectname\ was to 
keep every part of it as configurable as possible. This includes the core
framework, the plugins, the simulation and the filter process.
Each plugin can specify its own settings in a separate section of the main
\insertprojectname\ configuration file. The filtering mechanisms and the 
simulation part are configured in separate files. This gives the analyst 
the possibility to optimize the framework depending on individual needs and
thereby determine the verbosity of the results.

\section{The Core Framework}
\label{sec:core}
\begin{figure}
    \centering
    \includegraphics[height=0.31\textheight]{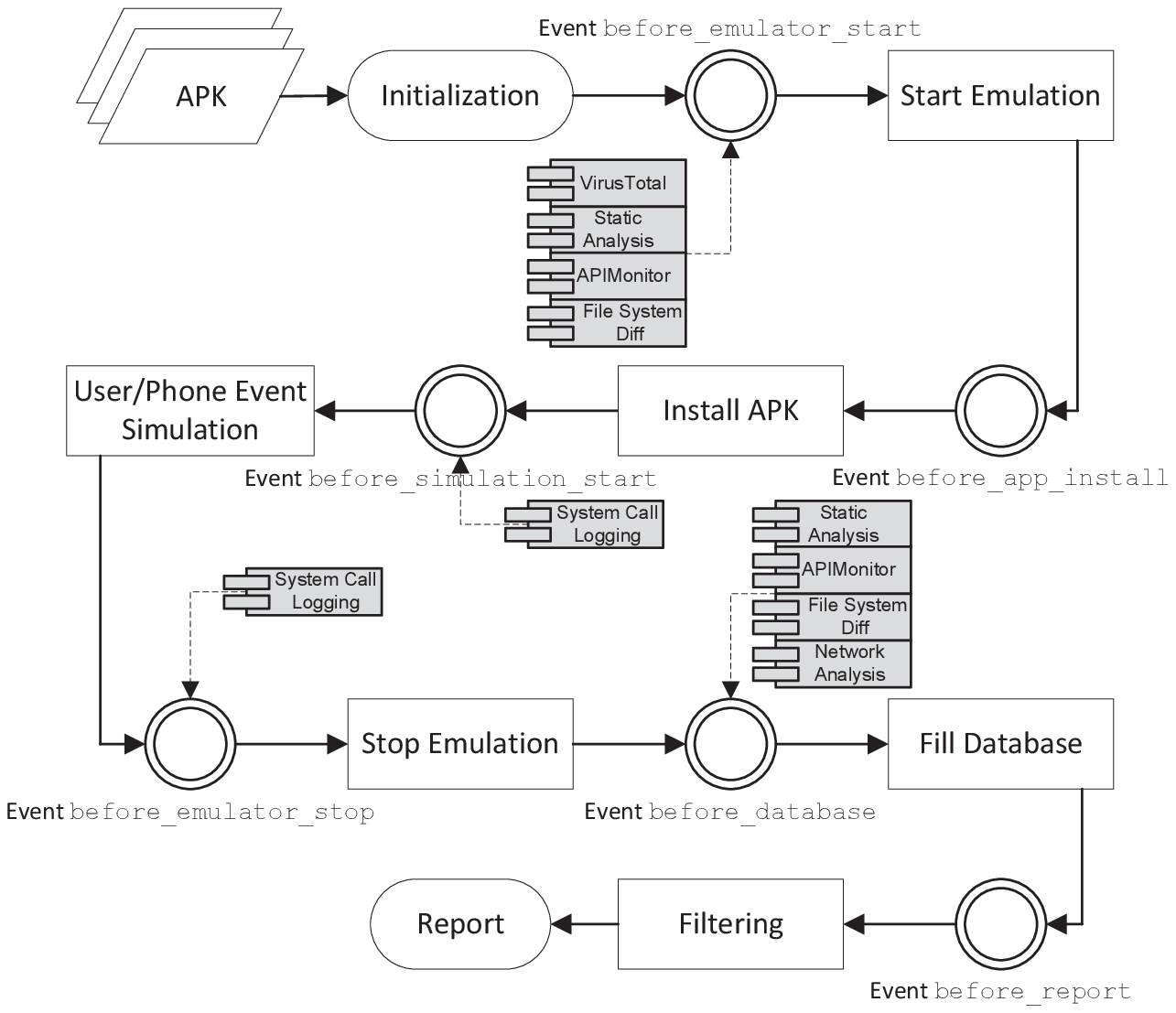}
    \caption{\insertprojectname\ workflow}
    \label{fig:workflow}
\end{figure}

In this section, the core of \insertprojectname\ is described by outlining the 
workflow of an analysis run. Figure \ref{fig:workflow} provides a high level overview of
the workflow.

\subsection{Initialization}

The first thing \insertprojectname\ does at startup is to load 
its default configuration, which is appropriate for most use cases.
If a custom configuration file is provided, these settings overrule the
default configuration.
Applications may behave differently on different Android versions and also
security controls can change from version to version. Therefore, an important
configuration setting for the dynamic analysis is the Android version
(\insertprojectname\ currently supports Android versions 2.3, 4.0, 4.1 and 4.2).

Then, \insertprojectname\ creates a clean environment for emulation and simulation purposes. 
The emulator is started with a temporary image, which is copied from a clean source.
This approach allows an easy comparison between the possibly infected 
and the clean image/filesystem.

As the last initialization step, the framework creates a unique directory for
the analysis, where all data (e.g. configurations, results, 
screenshots, ...) generated during the analysis is stored. All plugins
should save their results to this directory.

\subsection{Load Plugins}
The next step for the framework is to load the plugins. To ensure 
that an error in one plugin does not affect the overall analysis, 
plugins are being disabled for the current analysis run if any 
kind of error occurs.

Plugins can register on predefined events on which their methods 
get called by the framework. Each plugin can register methods 
to any event with a certain priority. Plugins with higher priorities 
get called first. 
The framework currently specifies the following events: 
\begin{itemize}
\item \textit{before\_emulator\_start, 
\item before\_app\_install, 
\item before\_simulation\_start,
\item before\_emulator\_stop,
\item before\_database} and 
\item \textit{before\_report}.
\end{itemize}
These events occur at different steps during the dynamic analysis and therefore
offer different perspectives on the analysis process and the analyzed app 
(have a look at section \ref{sec:plugins} for further details).

\subsection{Call Plugins: Part 1}

After the loading procedure, the plugins registered to the 
first event \textit{before\_emulator\_start} are called.
Then, the emulator is started with a temporary image for the configured 
Android version. By raising the \textit{before\_app\_install} event, plugins
that need to do work before the app is installed, are invoked. Then, the
app that should be analyzed is installed on the emulator. After another
event named \textit{before\_simulation\_start} and subsequent calling of registered
plugins, the simulation phase is starting.

\subsection{User \& Phone Event Simulation}


User and phone event simulation is one of the most important components of
the \insertprojectname\ framework, because malicious behaviour of most Android
malware is triggerd by certain user or phone events (as shown in section \ref{sec:architecture}
and \cite{malwaregenomepaper}). 

To allow a configurable simulation, a scripting language,
consisting of consecutive command and no control structures, 
is parsed by the framework. It makes heavy use of the Android Debug Bridge (ADB) Tool 
and the telnet interface of the Android emulator for executing its tasks.
The following simulation actions are currently available:
\begin{itemize}
    \item Starting and stopping installed Android applications,
    \item simulation of incoming and outgoing calls and text messages and changes to GSM state,
    \item changes to the smartphone's battery state,
    \item simulation of the emulator's location,
    \item user input using the \textit{monkey} program shipped with the Android SDK,
    \item unlocking and locking of the screen and
    \item execution of ADB commands and shell commands on the emulated system.
\end{itemize}

Commands within this script are for example \textit{callfrom '+431234567'} to simulate an incoming call from the specified 
number or \textit{changeLocation 'x-coordinate' 'y-coordinate'} to 
change the geo location of the emulated smartphone. An example for these commands
can be found in figure \ref{fig:simscript}.

\begin{figure}
\begin{lstlisting}
unlockscreen
sleep 3
startservices
startapp
screenshot
monkey 500
smsfrom "+49123456789" "Hi there."
callfrom "+49123456789"
setBatteryCapacity 5
setBatteryPowerState CHARGING
changeLocation "65.966667" "-18.533333"
\end{lstlisting}
\caption{An example simulation script.}
\label{fig:simscript}
\end{figure}


With the help of the command \textit{screenshot}, a screenshot 
of the current emulator's screen can be saved at any time. 
\insertprojectname\ ships with some 
predefined simulation scripts that should reveal most of the 
application's behaviour and are adjustable to specific needs.

\subsection{Call Plugins: Part 2}

After the simulation script reaches its end, the event \textit{before\_emulator\_stop} is invoked. 
This is the last chance for plugins to interact with the emulator, as after calling all plugins that are 
registered for this event, the emulator is stopped. For example, plugins can pull
files needed from the emulator's filesystem for later processing.  

If a plugin needs to process its results in any way before they are finally saved to the database, 
it can register for the next event named \textit{before\_database}. It is called directly before 
the final results of the plugins are written to the database.

Finally, the event named \textit{before\_report} is called, giving the
plugins a last chance to do any processing or cleanup work. 
The report is solely generated from the results that are saved in 
the database, so any processing of results that don't occur on 
database level won't have any effect on the report.

\subsection{Filtering \& Report Generation}

Before the report is generated, each plugin has the possibility to apply filters. 
Filters are used to determine which of the results that are saved in the database 
are worked into the report. As some plugins might generate a lot of log data, the
accuracy of the filters is crucial for the quality of the report and its usefulness 
to the analyst.

Filtering follows a blacklist approach, whereas certain values can 
be excluded from the blacklist using a separate whitelist, that
is only applied on blacklisted entries.
For example, in the case of a system call log, this allows blacklisting each \textit{open}
system call on paths starting with \textit{/data/} but excluding anything in the  
\textit{/data/data/} directory from blacklisting and thereby including it in the report.
Everything that doesn't show up in the blacklist is considered valuable and 
will be passed to the report generation.

Filters can be based on regular expressions or substring matches. For fine grained
filtering, filters can also be written in Python.

As the last step of the analysis process, the \insertprojectname\ framework generates 
the XML report based on the filtered results of the plugins. The report also
contains additional information such as the APK's filenames, hashes,
timestamps and information about plugins used. A shortened example of such
a report can be seen in figure \ref{fig:xmlreport}.

\begin{figure}
\begin{lstlisting}[language=xml]
<analysis>
  <filename>test.apk</filename>
  <hashes><hash type="md5">d331c96...</hash></hashes>
  <package>ananas.analysis</package>
  <starttime>...</starttime>
  <endtime>...</endtime>
  <plugins>
    <plugin enabled="True" name="lkm" state="loaded" />
    ...
  </plugins>
  <errors>...</errors>
  <results>
    <!-- plugin results -->
    <screenshots>...</screenshots>
    <virustotal>...</virustotal>
    <static>...</static>
    <apimonitor>...</apimonitor>
    <filesystemdiff>...</filesystemdiff>
    <syscalltrace>...</syscalltrace>
  </results>
</analysis>
\end{lstlisting}
\caption{Shortened example report.}
\label{fig:xmlreport}
\end{figure}

\section{Plugins}
	\label{sec:plugins}
In order to turn \insertprojectname\ into a useful analysis tool, plugins for various analysis methods 
were developed. 
They represent well-known techniques for static and dynamic malware analysis. 
Some of them were solely developed by the \insertprojectname\ development team. 
In order not to reinvent the wheel, several plugins were created that leverage existing 
tools and act as a wrapper.
The following plugins were developed for \insertprojectname:
\begin{itemize}
    \item Filesystemdiff
    \item Network analysis
    \item Systemcall logging
\end{itemize}
The following plugins wrap existing tools:
\begin{itemize}
    \item APKIL/APIMonitor
    \item Static analysis
    \item VirusTotal query
\end{itemize}

In the following sections, each plugin is described shortly.

\subsection{File System Diff}
The goal of File System Diff is to show changes in the
emulator's filesystem which might be caused by the analyzed Android application. It 
does so by comparing the emulator's filesystem before the emulator boots and after it has been powered off. 
The plugin registers to the hooks 
\textit{before\_emulator\_start} and \textit{before\_database}, where it mounts the emulator image into the analysis host's filesystem and creates a list of directories and files including their hash values. These lists are then compared to detect created, modified and deleted files. Files that only existed temporarily or were changed during the execution
of the emulator cannot be found by using this method.

In \insertprojectname, we try to avoid the usage of tools that require
root privileges. Thus, the ext4 filesystem is used for emulator images so, that they can be mounted with a slightly modified version
of \textit{ext4fuse} in userspace as an unprivileged user.

\subsection{Network Analysis}
\label{subsec:network}
Network traffic that occurs during the analysis is saved 
in a pcap file. Connections are extracted from the stored network traffic and provided in a readable format for later use in the report. For simplicity, in case of TCP and UDP traffic a connection is identified as a tuple 
of two IP address and port combinations. Traffic caused by other protocols is identified by the two IP addresses and the 
protocol used. In addition, IP packets for each connection are counted to give an overview of the total amount of network traffic.

To get a high-level view, two application level protocols are further examined: DNS and HTTP.
Therefore, queries made to DNS servers are extracted and the type and content 
of the query is shown. Furthermore, HTTP traffic is identified and relevant details such as URLs, header fields and parameters in GET and POST requests are extracted.

\subsection{Systemcall Logging/LKM}
\label{subsec:syscalls}
To provide a low-level view of the interaction between the analyzed application and
the Android system, a loadable kernel module is used to hook several system calls
and log their usage.
It allows \insertprojectname\ to trace all system calls that have been executed on the 
emulated system during the analysis. This is important to detect inter process communication (IPC), 
which is heavily used in Android. One example of IPC is the handling
of the Short Message Service (SMS). SMS messages received by the system are broadcasted to
applications (including the built-in SMS application), which requested
to receive this type of IPC message.
A trick that is used by malicious applications is to instrument a web browser via 
IPC to open a certain URL and exfiltrate data. 
This is why the logged system calls cannot be restricted to the analyzed application. 
Instead, all system calls on the whole system must be considered.

For each system call, the name and the parameters of the system call and its return value are logged. If an argument to a system call is a pointer to a string, the string is recorded.
The plugin includes hooks for mostly filesystem related system calls, such as system calls needed to
create, read, write, remove files and change permissions and ownership of files.
System calls used for changing user and group IDs are also logged.

Since the logging of system calls for all applications leads to very 
big log files, a good filtering mechanism must be put in place. 
The filters were written and improved by comparing 
the analysis results of several malicious and non-malicious 
applications.
System calls that represent normal
behaviour on an Android system are filtered because they do not provide valuable insight for 
the analyst.
Some popular root exploits for the Android system were 
whitelisted in order to make sure they show up in the report.
\insertprojectname's LKM plugin doesn't try to detect 
these exploits actively, but by whitelisting the exploits it 
is ensured that they are not accidentally blacklisted.

\subsection{APKIL/APIMonitor}
APIMonitor/APKIL \cite{apimonitor} was developed by Kun Yang at 
the Honeynet Project during the Google Summer of Code 2012 as 
an improvement for DroidBox \cite{droidbox}. By interposing APIs 
in the APK file and inserting monitoring code into the APK, API 
call logs can be retrieved.

The plugin registers to the event \textit{before\_emulator\_start} 
and uses APKIL to modify the APK before it is 
installed to the emulator. 
It is important to notice that all other plugins, such as the 
static analysis plugin, which are in some way dependent on the 
APK file, still use the unmodified version. The modified APK then logs its API calls to logcat, Android's  
logging system.
As the framework redirects the logcat files to the local
result directory on the analysis host, the plugin doesn't have 
to retrieve the logcat files manually. 
However, it registers to the hook \textit{before\_database} to 
process the logcat files and search for relevant APKIL entries.

\subsection{Static Analysis}
\label{subsec:static}
When analyzing applications, static analysis can be a very useful tool to gather 
information about the application. Especially the file \textit{AndroidManifest.xml},
which is part of every APK, is a precious source containing the application's 
permissions, activities, services, content providers, broadcast receivers as well as 
other useful information like the package name.

\insertprojectname' static analysis plugin's capabilities are to analyze the  
\textit{AndroidManifest.xml} and extract the application's permissions, services, receivers
and the package name.
As the application's manifest is compressed, apktool is used to convert the manifest to a readable format before 
extracting the information wanted. Apktool is a tool, which can 
decode resources within an APK and 
disassemble the compiled dex files to smali \cite{apktool}.

The static analysis plugin extracts strings from the disassembled dex files. These strings are then scanned for URLs. The knowledge, which URLs an Android application contains and probably connects to, 
can be a valuable information, since a lot of malicious Android applications try to send
personal information from the users' smartphones to external servers or 
try to contact their C\&C servers.

The next step of the static analysis plugin is to check each 
resource within the APK for its 
filetype. The purpose of this is to identify binaries or native libraries that ship with the APK 
to be executed later and might contain exploits. For malware analysts, 
it might be suspicious if an application contains a file called \textit{image.png}, which is 
detected as an executable.

To do its work, this plugin registers to the event
\textit{before\_emulator\_start} and starts the static analysis
in a background thread in order not to delay the dynamic analysis.
The plugin waits for the background thread to finish on the
\textit{before\_database} event. 

Although obfuscation and similar techniques can limit the benefit of static analysis for 
malware detection, it is still a valuable source of information, especially when it comes to
analyzing Android applications. As long as the specification of Android applications doesn't 
change, the \textit{AndroidManifest.xml} will always contain easily extractable and valuable 
information about the APK.

\subsection{VirusTotal Query}
\label{subsec:virustotal}
VirusTotal is a platform that allows the user to submit a file or the hash
of a file and receive the results of several antivirus scan engines.

This plugin provides the possibility to automatically include the results of 
VirusTotal into the final report. It does so by searching for the APK's hash on
VirusTotal using the hook \textit{before\_emulator\_start}. 
If VirusTotal isn't aware of the APK's hash, the plugin also provides the 
possibility to upload the APK and fetches the results of the analysis later.

\section{Experimental Evaluation}
\label{sec:evaluation}

The evaluation of \insertprojectname\ was done based on samples within the 
\textit{Android Malware Genome Project} (AMGP) \cite{malwaregenome}.
The whole set of 1,260 samples was analyzed using the framework to evaluate the 
robustness of the system. \insertprojectname\ proved to be capable of analyzing
this amount of samples without showing any fatal errors that would lead to a crash
of the framework or a stop of an analysis run.

For the first experimental evaluation of the framework's detection capabilities,
we analyzed a small subset of six randomly chosen samples from within the \textit{AMGP} 
supplemented by two more recent samples. The resulting reports were manually evaluated.
A more extensive evaluation on a larger amount of samples has to be conducted in the future.

In this section, we first present the results of this experimental evaluation. Problems 
and challenges faced during the evaluation are discussed subsequently. Finally, we 
draw a comparison between \insertprojectname\ and Andrubis \cite{andrubispost}
regarding the tested subset of samples within the \textit{AMGP}.

\subsection{Observation of Malicious Behaviour}

The \textit{AMGP} categorizes the samples by their payload functionality.
The categories are \textit{privilege escalation}, \textit{remote control}, \textit{financial charges}, 
and \textit{personal information stealing} \cite{malwaregenomepaper}.
Apart from \textit{remote control}, two samples of each category were analyzed 
with \insertprojectname. The \textit{remote control} category was excluded due to
the absence of reachable command \& control servers for the samples within the \textit{AMGP}
at the time of doing the evaluation. Additionally, two more recent samples were analyzed
(\textit{Android/SystemSecurity.A} and \textit{Trojan:Android/Maistealer.A}).

Each sample was analyzed with \insertprojectname\ using an Android 4.1 based emulation environment.
All plugins were enabled and an extensive user and phone event simulation script was used. 
The results of the experimental evaluation are shown in table \ref{tab:detection}.
The detection of malicious behaviour results in a \textit{Yes} for the respective category. 
If parts of the malicious behaviour or at least an indicator for such can be found in the report, 
this is described with \textit{Partly}. A detection failure results in a \textit{No}.
A hyphen (\textit{-}) symbolizes that the sample contains no malicious payload for the respective category.

As an example for the evaluation process, we discuss the analysis and results of 
two samples (DogWars and GGTracker) in greater detail below.

\begin{table*}[t]
    \centering
    \begin{tabular}{l c c c c c c c}
        \toprule
Application & Privilege Escalation & \multicolumn{3}{c}{Financial Charges} & \multicolumn{3}{c}{Personal Information Stealing} \\
\cmidrule(r){3-8}
             &        & Phone Call & SMS     & Block SMS & SMS    & Phone Number & User Account \\
\midrule
GGTracker    & -      & -          & No/No   & No/No     & Yes/No & Yes/Yes      & -            \\  
DogWars      & -      & -          & Yes/Yes & -         & -      & -            & -            \\
Gingermaster & Partly/Partly & -   & -       & -         & -      & Yes/No       & -            \\
AnserverBot  & -      & -          & -       & -         & No/No  & -            & -            \\
SndApps      & -      & -          & -       & -         & -      & -            & No/No        \\
DroidKungFu  & Partly/No & -       & -       & -         & -      & Yes/No       & -            \\ 
Pjapps       & -      & -          & No/No   & No/No     & -      & Yes/Yes      & -            \\
Trojan:Android/SystemSecurity.A*
             & -      & -          & No/No  & No/No     & -      & Yes/Yes      & -            \\
Trojan:Android/Maistealer.A*
             & -      & -          & -      & -         & No/No  & Partly/Partly & -            \\
    \end{tabular}
    \caption{Observation of malicious behaviour possible in \insertprojectname\ / in Andrubis. 
        Samples marked with \textit{*} are not part of the malware genome project.}
    \label{tab:detection}
\end{table*}

\subsubsection{Android.Dogowar}
Android.Dogowar (DogWars) is a trojan horse, which sends text messages to all contacts saved on the device.
It also sends text messages to a hardcoded number. This malware is a repackaged 
version of a legit game called \textit{Dog Wars}. \cite{dogwaranalysis}

The AMGP categorizes this sample's malicious payload into the category 
Financial Charges (subcategory SMS). This means, that the sample charges the
user financially by sending text messages.
\insertprojectname\ can detect this malicious behaviour. More precisely, the APKIL
plugin reveals the actions of this trojan. The first action of DogWars is to 
query the contacts application that ships with Android for contacts stored
on the device. Each contact for which a phone number exists will be forwarded a 
text message with the content
\textit{I take pleasure in hurting small animals, just thought you should know that}. 
Also, a text message to the number 73822 with the content \textit{text} is sent. All
this action occurs without the user's knowledge, leaving the user with the bill
for the sent SMS.

The network analysis plugin shows that \texttt{POST} and \texttt{CONNECT} HTTP requests to the 
URL \texttt{\url{http://kagegames.com/dw/process_cmds3.php}} are sent. 
The network analysis plugin also shows that no personal data left the phone
through these connections. As this malware is a repackaged version of a legit 
game, these connections might be part of the original game's behaviour.

\subsubsection{GGTracker}
The analyzed version of GGTracker hides inside a battery management
application and its main functionality is to sign-up the phone to premium 
SMS subscription services \cite{ggtrackeranalysis}. The AMGP classifies the 
malicious payload of GGTracker into Financial 
Charges (subcategory SMS) and Personal Information Stealing (subcategories SMS and Phone Number).

\insertprojectname\ is able to detect the information stealing part of the
payload. It is not possible for \insertprojectname\ to detect the 
financial charges that occur via premium SMS subscription services, as
the server used for subscribing to these services wasn't reachable
anymore at the time of writing this paper.

A first analysis of the sample showed that DNS queries for 
\texttt{\url{ggtrack.org}} and \texttt{\url{www.amaz0n-cloud.com}} failed. For the purpose of
generating better results, the emulator's hosts file was modified
to redirect these domains to a local server, where netcat was used as a sink for
incoming connections to port 80. This enabled us to record the HTTP requests 
sent by GGTracker and therefore get better analysis results.

The static analysis plugin showed the two URLs,
\texttt{\url{http://ggtrack.org/SM1c?device_id=}} and
\texttt{\url{http://www.amaz0n-cloud.com/droid/droid.php}}, the sample connected 
to. The APKIL plugin revealed that the sample requested access to the
SMS database several times but couldn't show that the SMS would leave 
the phone.
The File System Diff plugin showed that the following 
three files were added by the sample in its directory in 
\texttt{/data/data/t4t.power.management/}:
\begin{itemize}
\item \texttt{shared\_prefs/carrier.xml},
\item \texttt{shared\_prefs/t4t.power.management\ \_preferences.xml} and
\item \texttt{shared\_prefs/phone.xml}.
\end{itemize}
The names of the files \texttt{carrier.xml} and \texttt{phone.xml} alone
might raise concerns about sensitive data, which could be stored in 
these files.

The most beneficial plugin for the analysis of this sample is the 
network analysis. It clearly shows that the emulator's phone number is 
sent to \texttt{\url{ggtrack.org/SM1c}} via a \texttt{GET} request multiple times.
It also shows that the simulated incoming SMS that have been received
are forwarded to \texttt{\url{www.amaz0n-cloud.com/droid/droid.php}}. This is
done via a \texttt{POST} request, which contains the phone number of the
receiving as well as the phone number of the sending device, the carrier,
the content of the message, and the version of Android that is currently 
running on the receiving device.

\subsection{Challenges During the Evaluation}
The biggest challenge during the evaluation was that most command \& control
servers for older malware samples were not reachable anymore. 
This was especially problematic with personal information stealing samples.
We can often see that the data is accessed by the analyzed sample, but
because the command \& control server is not reachable the data is never
sent. By using DNS redirection and a netcat listener as a sink for HTTP 
requests from GGTracker, we were at least able to receive the requests of
this malware sample.

The unreachability of C\&C servers is also a problem with malware that fetches and loads code
at runtime and carries no malicious payload itself.
Some malware acts only as an installer for other Android applications. The 
actual malicious payload is contained in the application to be installed. 
While an application can dynamically load code, it cannot install other 
applications without prompting the user for permission.
\insertprojectname\ currently does not simulate a user approving an application installation and
therefore the actual malicious payload may never be installed and executed.

Another challenge for the dynamic analysis is to trigger every malicious behaviour 
of an application. While \insertprojectname\ ships with some predefined simulation
scripts which reveal most of the malicious activity, there is still behaviour that is not 
triggered (e.g. blocking of an incoming text message from a phone number that is 
specified in the application's code).

The usage of APIMonitor can also be a problem during the analysis. 
The AnserverBot malware family for example checks itself for integrity before the malicious 
payload is executed, as described by Zhou and Jiang in \cite{anserverbotanalysis}.
The sample is not activated at all if APIMonitor is used. With a disabled APIMonitor plugin
some activity can be observed, although the described malicious behaviour is still
not triggered.



\subsection{Comparison of \insertprojectname\ to Andrubis} 
To compare the quality of the reports generated by \insertprojectname\ to the
reports that have been generated by Andrubis, the test samples have 
been uploaded to Andrubis. It is worth noting, that each sample 
has already been analyzed by Andrubis prior to this evaluation. If C\&C servers
were still reachable at that time, results may be different.
The reports of Andrubis and \insertprojectname\ were compared manually.
As \insertprojectname\ focuses on dynamic analysis, only the dynamic parts
of the reports are compared against each other. However, it is obvious that the 
static analysis of Andrubis is superior to the static analysis plugin of \insertprojectname.

Unfortunately, no detailed description of the inner workings of the Andrubis 
system could be found. Therefore, it is not always clear why and how some behaviour
can or can not be detected with Andrubis.
We suppose that the differences in the detection of malicious behaviour between 
\insertprojectname\ and Andrubis mostly result from the use of different analysis techniques. 
Also, differences in the implemented detection evasion techniques or user and phone event simulation 
may be responsible for the different results.

Table \ref{tab:detection} shows the results of the comparative tests with Andrubis.
It is clear that both \insertprojectname\ and 
Andrubis are not able to detect every malicious behaviour with their dynamic analysis
modules. The challenges for the dynamic analysis that lead to detection failures might
be similar for both frameworks.

\section{Conclusion \& Future Work}
In this paper, \insertprojectname, an extendable framework for the analysis
of Android applications, as well as its plugins have been introduced. The 
main contributions of this paper are:
\begin{itemize}
\item \insertprojectname, which is a ready to use framework for dynamic
and static analysis of Android applications. It is highly configurable and 
independent of the emulated Android version.
\item An abstraction layer for simple user interaction and phone event simulation, 
which is easy to adjust.
\item A clean and defined interface for plugin developers to facilitate the
development of new plugins, regardless of whether they are intended for static or
dynamic analysis.
\item Six plugins for static and dynamic analysis of Android applications,
whereby three of them were developed solely for \insertprojectname\ and three are 
wrappers around existing tools.
\end{itemize}

The evaluation showed that it is important to simulate different phone events and 
user input during the dynamic analysis, since otherwise certain behaviour of the
application cannot be triggered.

It also showed that none of the implemented analysis methods was capable of 
detecting every malicious behaviour. Therefore, it is not sufficient to build a tool 
providing a single analysis method. It is important to have an expandable framework 
in which new or improved analysis methods can be integrated as plugins.
\insertprojectname\ aims to be such a framework.

By analyzing all samples within the Android Malware Genome Project, 
\insertprojectname\ proved to be a robust framework.
The developed plugins also proved to be effective by recognizing most of the 
application's malicious behaviour.

Possible improvements of the \insertprojectname\ framework, which are left for 
future work, are to evade detection of the analysis environment by malware or the
improvement of plugins. To evade the detection of the Android emulator,
the IMSI and IMEI of the emulator could be altered, but effective detection evasion
still needs more research. 
It is also a possibility to create new plugins for \insertprojectname\ or adopt 
existing tools for the analysis of Android applications as they are released.

Another topic, which needs more attention, is the development of effective filters 
for the plugins. The problem that has to be solved is to make sure that all relevant 
output of a plugin that can be mapped to malicious behaviour is shown in the report, 
while also making sure that non-relevant output does not show up. As some plugins,
e.g. the \textit{LKM} plugin, are very verbose, it is tempting to blacklist most 
of the output to make the final report smaller. The possibility to accidently blacklist 
relevant output is thereby tremendous.

One problem that occurs during the analysis, especially of older malware samples, is
that the C\&C servers are not reachable anymore. A simple simulation of such
C\&C servers could improve the generated results a lot.

Since the Android system is still quite young compared to other popular
operating systems, a lot of work has to be done to address security concerns.
The analysis of potentially harmful applications is one important aspect in this 
field to which this paper contributes.

\section*{Acknowledgements}
\label{sec:acks}
We would like to thank Christian Nösterer and Thomas Traunmüller for contributing
to \insertprojectname\ during their time in Hagenberg.

We would also like to thank Martin Brunner for his guidance in the early stages of
this project and Daniel Bäumges, whose \textit{TaintDroid Runner} project served as
initial base for \insertprojectname.

\IEEEtriggeratref{15}



\bibliographystyle{IEEEtran}


\bibliography{IEEEabrv,ananas-final-paper}


\end{document}